\documentclass[pra,aps,twocolumn,showpacs]{revtex4}
\usepackage{bm}
\usepackage{amsmath}
\usepackage{amsfonts}
\usepackage{amssymb}
\usepackage{amstext}
\usepackage{graphicx}
\usepackage{dcolumn}
\newcommand{\bd}{\begin{document}}
\newcommand{\ed}{\end{document}}
\newcommand{\bc}{\begin{center}}
\newcommand{\ec}{\end{center}}
\newcommand{\bfr}{\begin{flushright}}
\newcommand{\efr}{\end{flushright}}
\newcommand{\lt}{\left}
\newcommand{\rt}{\right}
\newcommand{\vs}{\vspace}
\newcommand{\hs}{\hspace}
\newcommand{\beq}{\begin{equation}}
\newcommand{\eeq}{\end{equation}}
\newcommand{\lb}{\linebreak}
\newcommand{\pb}{\pagebreak}
\newcommand{\mb}{\makebox}
\newcommand{\fb}{\framebox}
\newcommand{\mc}{\multicolumn}
\newcommand{\ben}{\begin{enumerate}}
\newcommand{\een}{\end{enumerate}}
\newcommand{\bit}{\begin{itemize}}
\newcommand{\eit}{\end{itemize}}
\newcommand{\un}{\underline}
\newcommand{\lefq}{\lefteqn}
\newcommand{\ba}{\begin{array}}
\newcommand{\ea}{\end{array}}
\newcommand{\beqa}{\begin{eqnarray}}
\newcommand{\eeqa}{\end{eqnarray}}
\newcommand{\beqas}{\begin{eqnarray*}}
\newcommand{\eeqas}{\end{eqnarray*}}
\newcommand{\bfg}{\begin{figure}}
\newcommand{\efg}{\end{figure}}
\newcommand{\bds}{\begin{displaymath}}
\newcommand{\eds}{\end{displaymath}}
\newcommand{\btb}{\begin{tabbing}}
\newcommand{\etb}{\end{tabbing}}
\newcommand{\para}{\parallel}
\newcommand{\pad}{\partial}
\newcommand{\nn}{\nonumber}
\newcommand{\la}{\leftarrow}
\newcommand{\ra}{\rightarrow}
\newcommand{\lgla}{\longleftarrow}
\newcommand{\lgra}{\longrightarrow}
\newcommand{\La}{\Leftarrow}
\newcommand{\Ra}{\Rightarrow}
\newcommand{\Lra}{\Leftrightarrow}
\newcommand{\Lgla}{\Longleftarrow}
\newcommand{\Lgra}{\Longrightarrow}
\newcommand{\lan}{\langle}
\newcommand{\ran}{\rangle}
\renewcommand{\a}{\alpha}
\renewcommand{\b}{\beta}
\newcommand{\g}{\gamma}
\newcommand{\G}{\Gamma}
\renewcommand{\d}{\delta}
\newcommand{\eps}{\epsilon}
\newcommand{\Th}{\Theta}
\newcommand{\s}{\sigma}
\newcommand{\lam}{\lambda}
\newcommand{\D}{\Delta}
\newcommand{\vare}{\varepsilon}
\newcommand{\pr}{\prime}
\newcommand{\ro}{\rho}
\newcommand{\nab}{\nabla}
\newcommand{\m}{\mu}
\newcommand{\n}{\nu}
\newcommand{\Sg}{\Sigma}
\newcommand{\p}{\pi}
\newcommand{\R}{I\!\!R}
\newcommand{\om}{\omega}
\newcommand{\Om}{\Omega}
\newcommand{\ze}{\zeta}
\newcommand{\vart}{\vartheta}
\newcommand{\tri}{\triangle}
\newcommand{\f}{\frac}
\newcommand{\iny}{\infty}
\newcommand{\pro}{\propto}
\newcommand{\np}{\newpage}
\begin{document}

\title{Localization of Dirac-like excitations in graphene in the presence of
smooth inhomogeneous magnetic fields}
\author{Pratim Roy, Tarun Kanti Ghosh, Kaushik Bhattacharya}
\affiliation{Department of Physics, Indian Institute of Technology-Kanpur,
Kanpur-208 016, India}

\begin{abstract}
 
The present article discusses magnetic confinement of the
Dirac excitations in graphene in presence of inhomogeneous magnetic
fields. In the first case a
magnetic field directed along the $z$ axis whose magnitude is
proportional to $1/r$ is chosen. In the next case we choose a more
realistic magnetic field which does not blow up at the origin and
gradually fades away from the origin. The magnetic fields chosen do
not have any finite/infinite discontinuity for finite values of the
radial coordinate.  The novelty of the two magnetic fields is related
to the equations which are used to find the excited spectra of the
excitations. It turns out that the bound state solutions of the
two-dimensional hydrogen atom problem are related to the spectra of
graphene excitations in presence of the $1/r$ (inverse-radial)
magnetic field. For the other magnetic field profile one can use the
knowledge of the bound state spectrum of a two-dimensional cut-off
Coulomb potential to dictate the excitation spectra of
graphene. The spectrum of the graphene excitations in presence of the
inverse-radial magnetic field can be exactly solved while the other case
cannot be. In the later case we give the localized solutions of the
zero-energy states in graphene.

\end{abstract}


\date{\today}
\maketitle
\section{Introduction}
In recent years graphene, which is a single layer of carbon atoms on a
honeycomb lattice, has generated a lot of interest in the research
community \cite{novo,zheng,rmp}.  This is essentially because of the
experimental realization of graphene sheets in various laboratories
and its possible applications in nanoscale devices.  An electron in a
graphene layer is described by a massless two-dimensional (2D)
Dirac-like equation which produces a gapless linear energy spectrum
close to the Dirac points $K$ and $K^{\prime}$.  The Dirac electron
imitates various relativistic phenomena like Klein tunneling
\cite{klein}, unconventional Hall effect \cite{novo,zheng,hall} and
Zitterbewegung \cite{zb}.  To make electronic devices, the first thing
is to confine the electrons.  This can be attempted by either using an
electric field or a magnetic field. Although zero-energy bound states
or quasi-bound states may be created using electrostatic barriers or
external potentials, the Dirac electron can not be always confined
in this way due to its massless nature \cite{mutu}. Particularly, the
electrostatic barrier becomes completely transparent for normal
incidence of the electrons.

Some schemes have been proposed to tackle the problem of confining the
electrons. For example, A. de Martino et al., has given an interesting
proposal to confine the electron based on an inhomogeneous magnetic field
\cite{martino}.  Later on various inhomogeneous magnetic field
configurations have been used in an attempt to confine the electrons
or create bound states \cite{pereira}. Among the various field
configurations, exponentially decaying magnetic fields \cite{ghosh},
Gaussian magnetic field \cite{lee}, square well magnetic barrier
\cite{martino}, non-zero magnetic fields in a circular dot
\cite{wang}, magnetic fields corresponding to various solvable
potentials \cite{nieto} etc., have been considered.  In all these
studies, discontinuous and/or inhomogeneous (in some specific
direction) magnetic field profiles were considered.

Recently, Masir et al. \cite{mutu} has considered a finite-size
magnetic dot with a constant magnetic field ($B_z$) within a circular
region with radius $R$, otherwise zero i. e. $ {\bf B} = B_z
\Theta({\bf R} - {\bf r}) \hat z$.  It is found that bound state does
not exist in this configuration.  In the present article our objective
is to use different types of axial magnetic fields, which are
continuous for any finite radial distance, to examine the possibility
of creating bound states in graphene. In the first case we will
consider the inverse-radial field pointed along the negative $z$-axis
and whose magnitude is $\sim 1/r$. One can exactly solve for the
Dirac electron-like excitations in graphene in presence of such a
magnetic field. We will focus on the localized excitations in the
present case. One of the surprising outcomes of the above analysis is
an interesting relation between the bound state spectrum of the 2D
Hydrogen atom and the energy spectrum of the magnetically confined
graphene excitations. In this article it will be seen that the
Schr\"{o}dinger equation of the 2D Hydrogen atom plays a very
important role for determining one of the components of the Dirac-like
excitations of graphene in presence of the inverse-radial magnetic
field. Due to the complex nature of the Dirac equation it turns out
that the relevant Coulomb potential in the 2D Hydrogen atom problem
becomes angular momentum dependent. Except this complication of an
angular momentum dependent Coulomb potential the 2D Hydrogen atom
bound state spectra gives us a clue of the bound state energies of the
graphene excitations in an inverse-radial magnetic field.

Although for an inverse-radial magnetic field one can find the spectrum
of the localized massless excitations of graphene exactly, still the
magnetic field profile is not realistic near the origin.  Noting this
fact, in the second case, we try to see the properties of the
localized excitations in graphene in a more realistic magnetic field
by introducing a cut-off parameter (i.e, the magnetic field does not
blow up at the origin). As a consequence of introducing this cut-off
parameter the problem does not remain an exactly solvable
one. Nevertheless localized zero-energy states can be exactly
determined and it will be seen that in this case also the zero-energy
states are infinitely degenerate.

Before we list the materials of the present article in the last
paragraph it is useful to give a brief discussion about the magnetic
field profiles used in the article. The authors are aware of the fact
that the kinds of magnetic fields used for the calculations are not
generally used in laboratories. But in the future one may produce a
magnetic field which has a high but finite strength in the axial
direction and slowly the field strength falls as one moves outwards
from the axis. Using various cylindrical dipolar magnets, with the
magnetic axis along the cylindrical axis, with varying magnetic
strength one can in principle produce a field which can resemble the
realistic magnetic field with a cylindrical symmetry as assumed in
this article. The central solid cylindrical magnet with the highest
magnetic strength must be concentric with many annular cylindrical
magnets with lower and lower magnetic strength. The magnetic strengths
can be imparted suitably in the laboratories. With two such set ups
placed one above the other, such that the opposite poles face each
other one can produce magnetic fields of various cylindrical
symmetries in the intervening region.

The materials in this article are presented in the following manner. A
brief and concise scheme of the mathematical framework which will be
utilized throughout the article is presented in the next section.
The spectrum of the localized excitations of graphene in the presence
of an inverse-radial magnetic field is given in section \ref{coul}. In
section \ref{realb} we present the exact zero-energy solutions for the
graphene excitations in presence of a more realistic magnetic field
which does not diverge at the origin. The last section \ref{conclu}
concludes the article with a brief discussion and summarizes the
findings presented in the article.
\section{The Model}
\label{mod}
The quasi-particles in graphene follow a Dirac-like
Hamiltonian given by 
\beq 
H = v_F{\bm{\sigma} \cdot {\bf P}}\,,
\label{bham} 
\eeq 
where $v_F \approx 10^6 m/s$ is the Fermi velocity,
${\bm\sigma}=(\sigma_x,\sigma_y)$ are the Pauli matrices and
$\bm{P}=\bm{p}+{\bm A}$ is the canonical momentum with vector
potential ${\bf A} $. For the sake of simplicity as well as clarity we
will use the convention in which $\hbar=e=1$, where $e$ stands for the
magnitude of the electronic charge.  The eigenvalue equation for the
above Hamiltonian is $ H \psi(x,y) = E \psi(x,y) $ where $\psi(x,y)$
is a two-component spinor given by
\beq 
\psi = \left( \ba{c} \psi_1(x,y) \\ 
\psi_2(x,y) \ea \right)\,. 
\nonumber
\eeq
Here the two components $\psi_{1,2}(x,y)$ denote the electron amplitude on 
two sites in the unit cell of a honeycomb lattice. 
We now write the eigenvalue equation in the following coupled form 
as
\beq
\Pi_-\psi_2=\eps\psi_1\,,~~~~\Pi_+\psi_1=\eps\psi_2\,,
\label{eigen2} 
\eeq
where $\eps = E/v_F$ and $\Pi_{\pm}=P_x \pm iP_y$.  The general form of
the vector potential is of the form 
\beq 
A_x=y f(r)\,,~~~~A_y=- x f(r)\,, 
\eeq 
where the function $f(r)$ will be specified later. For this choice of
the vector potential the associated magnetic field is perpendicular to
the plane and it is given by 
\beq 
B_z=-2 f(r)- r f^\prime(r)\,.
\label{magnetic} 
\eeq 
Since the magnitude of the magnetic field only depends on $r$ it is
convenient to use the plane polar coordinates in the present case. In
plane polar coordinates one can write the component spinors as
\beq
\psi_1=r^{-{1}/{2}}e^{im\theta}\chi_1(r)\,,~~~~
\psi_2=r^{-{1}/{2}}e^{i(m+1)\theta}\chi_2(r)\,
\label{wf2} 
\eeq 
where $~m=0,\pm 1,\pm 2,\cdot \cdot \cdot$, stands for the angular
momentum quantum number of the pseudo-spinor solution of the Dirac
equation.  Using the above solutions and Eq.~(\ref{eigen2}) the
eigenvalue equation for $\chi_1$ can be written as
\beq
\left[-\f{d^2}{dr^2}+V^{\rm eff}_1(r)\right]
\chi_1=\eps^2\chi_1\,,
\label{chi11}
\eeq
where the effective potential term is 
\begin{eqnarray}
V^{\rm eff}_1(r)=V_1(r)+ \f{m^2-\f{1}{4}}{r^2}\,,
\label{veff1}
\end{eqnarray}
and
\begin{eqnarray}
V_1(r)= r^2f^2-2mf-2f - rf^\prime\,,
\end{eqnarray}
where the other term in the potential $V^{\rm eff}_1(r)$ is due to the
centrifugal barrier.  The factor of $1/4r^2$ appears because of the
presence of $r^{-{1}/{2}}$ in the solutions in Eq.~(\ref{wf2}).
Similarly for $\chi_2$ we obtain
\beqa
\left[-\f{d^2}{dr^2}+V^{\rm eff}_2(r)\right]\chi_2=\eps^2\chi_2\,,
\label{chi22}
\eeqa
where the effective potential is 
\begin{eqnarray}
V^{\rm eff}_2(r)=V_2(r)+ \f{(m+\f{1}{2})(m+\f{3}{2})}{r^2}\,,
\label{veff2}
\end{eqnarray}
and 
\begin{eqnarray}
V_2(r)=r^2f^2-2mf+rf^\prime\,.
\end{eqnarray}
It is easy to see that the above equations are standard radial
Schr\"odinger equations (in 2D) with the effective potentials
$V_{1,2}^{\rm eff} (r)$.  
\section{The inverse-radial magnetic field and the spectrum of the excitations}
\label{coul}
In this section we will concentrate on a magnetic field profile which
is given by
\begin{eqnarray}
B_z=-\frac{\lambda}{r}\,,
\label{cmag}
\end{eqnarray}
where $\lambda > 0$ is a constant with the dimension of inverse
length. In normal units the magnetic field will be represented as
$-{\hbar\lambda}/{er}$.  A field of the above type can be obtained by
choosing the $f(r)$ as
\beq
f(r)=\f{\lambda}{r}\,.
\label{fr1} 
\eeq
With the above form of $f(r)$ one can see that 
\beqa
V_1(r)= V_2(r)=-\f{\lambda(2m+1)}{r}+\lambda^2\,,
\label{cv1}
\eeqa 
\begin{figure}[ht]
\begin{center}\leavevmode
\includegraphics[width=95mm]{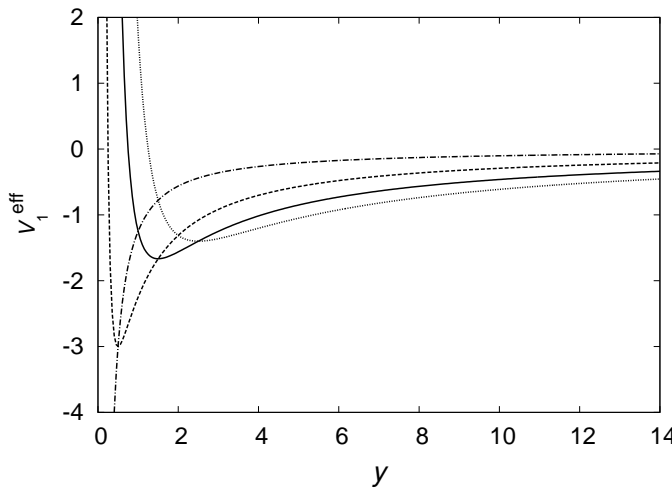}
\caption{Plots of the effective potential $V_1^{\rm eff} $ vs 
$y$ for $m = 0$ (dot-dashed), $m = 1$ (dashed), $m = 2$ (solid) and 
$m = 3 $ (short-dashed).}
\label{Fig1}
\end{center}
\end{figure}
\begin{figure}[h]
\begin{center}\leavevmode
\includegraphics[width=95mm]{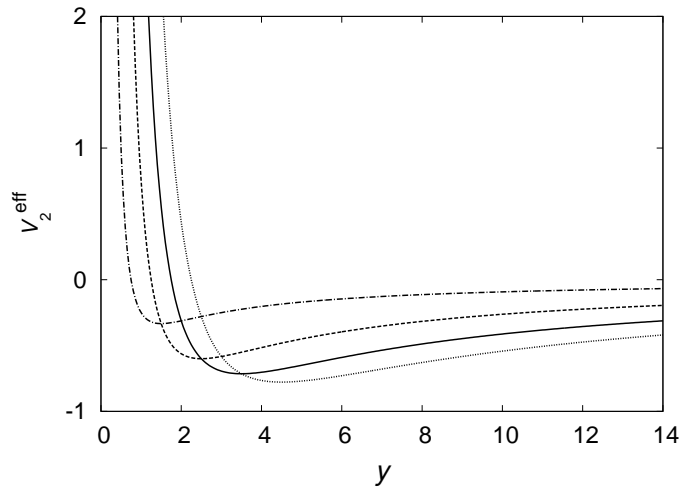}
\caption{Plots of the effective potential $V_2^{\rm eff} $ vs 
$y$ for $m = 0$ (dot-dashed), $m = 1$ (dashed), $m = 2$ (solid) and
$m = 3 $ (short-dashed).}
\label{Fig2}
\end{center}
\end{figure}
It is interesting to note that with the form of $f(r)$ as given in
Eq.~(\ref{fr1}), both the potentials $V_{1,2}(r)$ turns out to be of
the Coulomb-type for $m \ge 0$.  The shape of the effective potentials
$V^{\rm eff}_{1,2}(r)$ for some positive values of $m$ are plotted in
Fig.~\ref{Fig1} and Fig.~\ref{Fig2}. The shape of the effective
potentials clearly shows that there can be bound states for positive
values of the angular momentum quantum numbers.  For $m < 0$ the 
potentials $V_{1,2}(r)$ are not attractive and hence do not support any 
bound state. Consequently, for states with $m \ge 0$ the 
Dirac electron-like excitations in graphene due to the
inverse-radial magnetic field can be
understood by solving the radial Schr\"{o}dinger equation,
Eq.~(\ref{chi11}), in presence of a 2D Coulomb
potential. The novelty of the 2D Coulomb potential, appearing in the
Schr\"{o}dinger equations, in the present case is related to the fact
that the potentials depends on the angular momentum of the
excitations.  This observation is pivotal for the analysis which
follows.
\subsection{The 2D Hydrogen atom and the upper component of the
  graphene pseudospinor}
To find the ground state and excited state energies and the 
corresponding wave functions of the Dirac electron-like excitations 
in graphene in presence of an inhomogeneous magnetic field, as given in
Eq.(\ref{cmag}), we require to solve Eq.~(\ref{chi11}) where $V_1(r)$
in the effective potential is given in Eq.~(\ref{cv1}). One can write
Eq.~(\ref{chi11}) in terms of the function
\begin{eqnarray}
\phi_1(r)=r^{-1/2}\chi_1(r)\,,
\label{phichi1}
\end{eqnarray}
as
\begin{eqnarray}
-\frac{d^2 \phi_1}{dy^2} - \frac{1}{y}\frac{d\phi_1}{dy}+
\left[-\frac{2m+1}{y} + \frac{m^2}{y^2}\right]\phi_1=
\xi \phi_1\,,
\label{excited}
\end{eqnarray}
where $$y \equiv \lambda r\,,$$ is a dimensionless parameter and
\begin{eqnarray}
\lambda^2 \xi =\eps^2- \lambda^2\,.
\label{defl}
\end{eqnarray}
The energy eigenvalues are given by
\begin{eqnarray}
E=\pm \lambda v_F \sqrt{1 + \xi}\,.
\label{energy_eig}
\end{eqnarray}
Eq.~(\ref{excited}) can be converted to a confluent hypergeometric
equation in $x$ where
$$y=\frac{1}{\beta_N}x\,,\,\,\,\,\,\beta_N=\frac{2m+1}{N}\,,$$
for a real number $N$. Defining a new function $\Phi_1(x)$ as
$$\phi_1(x)=x^m e^{-x/2} \Phi_1(x)\,,$$ Eq.~(\ref{excited}) 
transforms into the following confluent hypergeometric differential equation:
\begin{eqnarray}
x\frac{d^2\Phi_1}{dx^2}+[(2m+1)-x]\frac{d\Phi_1}{dx}
-(-N+m+\frac12)\Phi_1=0\,.\nonumber\\
\label{meqn}
\end{eqnarray}
The most general solution of the above equation is
\begin{eqnarray}
\Phi_1(x)&=&A\,\,{_1F}_1(-N+m+\frac12, 2m+1, x)\nonumber\\
&+& B \,\,U(-N+m+\frac12, 2m+1, x)\,,
\label{gensol}
\end{eqnarray}
where ${_1F}_1(-N+m+\frac12, 2m+1, x)$ and $U(-N+m+\frac12, 2m+1,
x)$ forms a pair of linearly independent solutions of the confluent
hypergeometric equation. $A$ and $B$ are two arbitrary constants.

In our analysis of $\phi_1(x)$ the angular quantum number $m$ can be
zero or positive integers. The nature of $N$ will dictate the
properties of the confluent hypergeometric functions. Out of various
possibilities of choosing $N$ one is particularly interesting. If $N$
are half-integers as
\begin{eqnarray}
N=\frac12\,,\,\frac32\,,\,\frac52\,,\,\cdot\cdot\cdot\,,
\label{nval}
\end{eqnarray}
then $-N+m+\frac12$ is a negative integer or zero if we restrict the
$m$ values in an appropriate way. Defining
\begin{eqnarray}
n=N+\frac12=1,2,3,\cdot\cdot\cdot\,,
\label{snval}
\end{eqnarray}
the possible $m$ values for which $-N+m+\frac12$ can be negative or zero
turns out to be
\begin{eqnarray}
m=0,1,2,\cdot\cdot\cdot,n-1\,.
\label{smval}
\end{eqnarray}
When $-N+m+\frac12$ or $-n+m+1$ is a negative integer or zero then
${_1F}_1(-n+m+1, 2m+1, x)$ reduces to a finite polynomial in the
positive $x$ range. On the other hand if $-n+m+1$ is a negative
integer then $U(-n+m+1, 2m+1, x)$ and ${_1F}_1(-n+m+1, 2m+1, x)$ are
not linearly independent of each other and consequently the pair of
functions in Eq.~(\ref{gensol}) does not give the most general
solution of Eq.~(\ref{meqn}). In this special case one may think of
the second solution of Eq.~(\ref{meqn}) as $e^x\,U(m+n, 2m+1, -x)$
\cite{abromowitz}, which is linearly independent of ${_1F}_1(-n+m+1,
2m+1, x)$. But, the function $e^x\,U(m+n, 2m+1, -x)$ diverges at large
values of $x$ and consequently this second solution is also unsuitable
in the most general solution of Eq.~(\ref{meqn}).

If the second solution of the confluent hypergeometric equation is
omitted then the solution of Eq.~(\ref{meqn}) coincides with the
solution of the 2D hydrogen atom \cite{yang}. In this case
\begin{eqnarray}
\phi_1^{(n,m)}(y) & = & (\beta_n y)^m e^{-\beta_ny/2}\nonumber\\
&\times&\left[ {_1F}_1(-n+m+1, 2m+1, \beta_n y)\right]\,,
\label{phi1y}
\end{eqnarray}
up to a normalization constant and
\begin{eqnarray}
\beta_n=\frac{2(2m+1)}{2n-1}\,.
\end{eqnarray}
The eigenvalues in the present case given by
$\xi_{n,m}$ will be of the form
\begin{eqnarray}
\xi_{n,m}=-\frac{(2m+1)^2}{(2n-1)^2}\,,
\label{lambdan}
\end{eqnarray}
where $m=0, 1,2,3,\cdot\cdot\cdot$ and $n=1,2,3, \cdot\cdot\cdot$ the
positive integers only. From the expression of the energy eigenvalues
as given in Eq.~(\ref{energy_eig}) we get
\begin{eqnarray}
E_{n,m}=\pm \lambda v_F \sqrt{1 - \frac{(2m+1)^2}{(2n-1)^2}}\,.
\label{energy_eigv}
\end{eqnarray}
The negative energies in the above represent the holes in the valence band.
The exact energy eigenvalues of the Dirac electron-like
excitations in graphene in presence of an inhomogeneous magnetic field
is seen to depend on two quantum numbers. One of them is the radial
one, $n$, while the other is related to the angular momentum quantum
number $m$. 
\begin{figure}[h]
\begin{center}\leavevmode
\includegraphics[width=95mm]{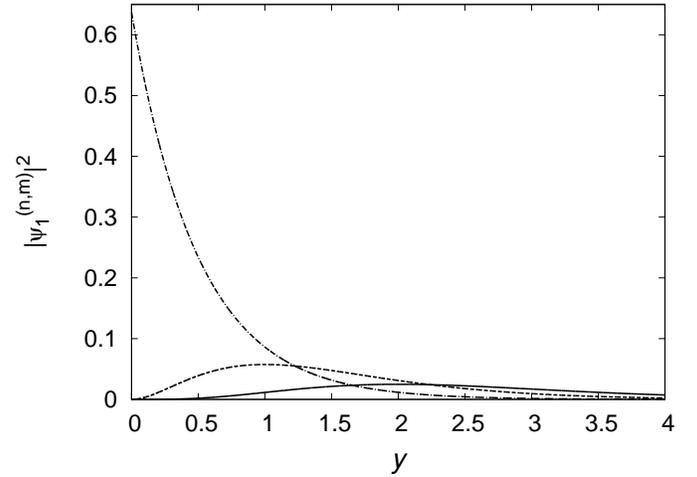}
\caption{Plots of the normalized probability density of the zero-energy
states vs $y$ for  $n=1,m = 0$ (dot-dashed),
$n=2,m = 1$ (dashed), and $n=3,m = 2$ (solid).}
\label{Fig3}
\end{center}
\end{figure}
%
\subsection{Degeneracies}
If one looks at the energy spectrum of the excitations one will see
that there are states which are degenerate.  The degeneracies can be
understood from the form of the energy eigenvalues as given in
Eq.~(\ref{energy_eig}).  It can be easily observed that
\beq
E_{1,0}=E_{2,1}=E_{3,2}=\cdot\cdot\cdot=E_{n,n-1}=\cdot\cdot\cdot=0\,.
\label{degvac}
\eeq
This implies that the ground states of the excitations are infinitely
degenerate. The corresponding wave function is given by
\beq
\psi^{(m+1,m)}(r,\theta) 
= N_{m} 
 \left(\ba{c} y^m e^{-y} e^{im\theta} \\ 0\\ \ea \right),
\eeq
where $ m = 0,1,2,\cdot\cdot\cdot$.  The normalization constant is
\beq
N_{m}= \sqrt{\frac{1}{2 \pi 2^{2(m+1)} \Gamma(2m +2)}}.
\eeq
The probability density plots for some of the zero-energy solutions
are given in Fig.~\ref{Fig3}. All of the curves show that the states
are localized near the origin and can be treated as proper bound states.

In general, there are many other infinitely degenerate states.
For example, the energy levels are infinitely degenerate if 
$(2m+1)/(2n-1) = p/q$, where $p$ and $q$ are odd integers with
$p\leq q $. The infinitely degenerate zero-energy state is obtained
for $ q = p$.
\subsection{The lower component of the pseudospinor}
\begin{figure}[h]
\begin{center}\leavevmode
\includegraphics[width=95mm]{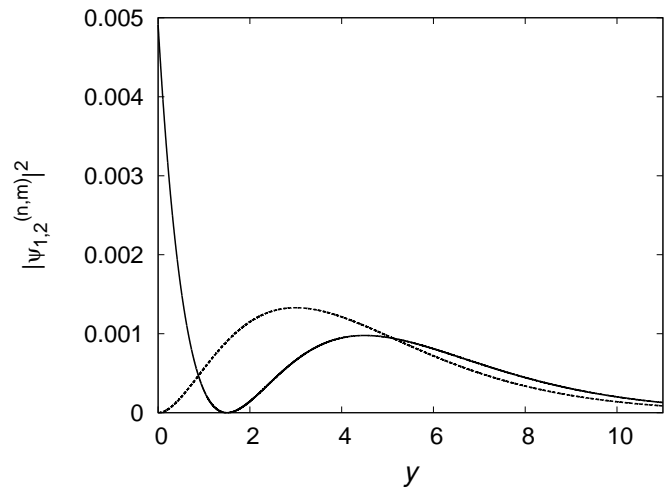}
\caption{Plots of the normalized density profile of the upper 
component (solid) and lower component (dash) for $n = 2, m = 0$.}
\label{Fig4}
\end{center}
\end{figure}

\begin{figure}[h]
\begin{center}\leavevmode
\includegraphics[width=95mm]{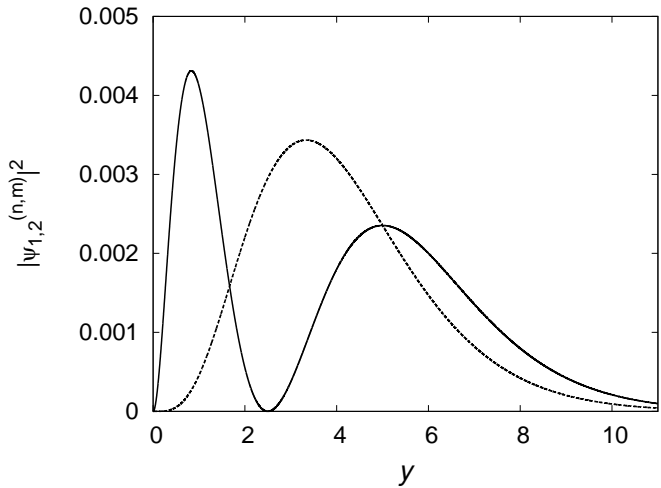}
\caption{Plots of the normalized density profile of the upper 
component (solid) and lower component (dash) for $n = 3, m = 1$.} 
\label{Fig5}
\end{center}
\end{figure}

For excited states, $\eps \ne 0$, the lower component of the
pseudospinor can be obtained by the relation
$\psi_2=\frac{1}{\eps}\Pi_+\psi_1$. In the present case $\psi_1(y)=
e^{im\theta}\phi_1(y)$ and consequently one can write
\begin{eqnarray}
\psi_2(y)=-\frac{i\lambda}{\eps}\,e^{i(m+1)\theta}\left[
\frac{d}{dy} - \frac{m}{y}+1\right]\phi_1(y)\,,
\label{psi2}
\end{eqnarray}
where $\phi_1(y)$ is as given in Eq.~(\ref{phi1y}). But the lower
component of the zero-energy states cannot be obtained from the
formula given above.  To obtain the the lower component of the
pseudospinor for the zero-energy case one can directly use the first
order equations in Eq.~(\ref{eigen2}). Solving the first order
equation $\Pi_-\psi_2= 0$ 
where $\Pi_-=e^{-i\theta}[-i\partial/\partial r - (1/r) \partial/\partial
\theta + i \lambda]$ and $\psi_2(r)=\phi_2(r) e^{i(m+1)\theta}$
one easily obtains  
\begin{eqnarray}
\phi_2(y)=\frac{e^y}{y^{m+1}}
\label{2ndsoln}
\end{eqnarray}
up to a normalization constant and where $y=\lambda r$. The solution
listed above is not square integrable and consequently it cannot be
taken as the lower component of the pseudospinors. Consequently the
lower component of the zero-energy states are all zero.  The modulus
squared pseudospinor components, $\psi_{1,2}^{(n,m)} $, for two 
specific cases as, $m=0,\,\,n=2$ and $m=1,\,\,n=3$, are plotted in 
Fig.~\ref{Fig4} and Fig.~\ref{Fig5}, respectively. 
Here $\psi_{1,2}^{(n,m)}$ stands for either the upper component
$\psi_{1}^{(n,m)}$ or the lower component $\psi_{2}^{(n,m)}$ of the
pseudospinor for arbitrary values of the quantum numbers $n$ and $m$.
Both the curves show that the probability densities of the excitations 
in both these cases are localized.
\subsection{Reversing the direction of the magnetic field}
It is interesting to see what happens to the spectrum of the excited
states when the magnetic field direction is reversed. From
Eq.~(\ref{fr1}) it can be seen that the magnetic field direction
reversal is equivalent to changing the sign of $\lambda$. When the
sign of $\lambda$ changes the potentials $V_{1,2}(r)$ do not remain
attractive for $m \ge 0$ and so localized states of the Dirac
excitations cannot be obtained. On the other hand for negative angular
momentum states, $m \le -1$, Eq.~(\ref{cv1}) reveals that an
attractive potential is possible when $\lambda<0$. Consequently, when
the magnetic field direction is reversed one obtains localized states
for the Dirac excitations for $m \le -1$. From the form of the
effective potentials in the Schr\"{o}dinger equations of $\chi_1$ and
$\chi_2$, as given in the initial part of section \ref{mod}, it can be
easily verified that if one changes the sign of $\lambda$ and replaces
$m$ by $-(m+1)$ (for a negative $m$ where $m \le -1$) then $\chi_1$ and
$\chi_2$ just gets interchanged. Consequently we can say that the
effect of the magnetic field reversal amounts to interchanging the
components of the pseudospinors previously obtained for the magnetic field
pointing in the negative $z$-axis  with their
positive $m$ values replaced by $-(m+1)$ where now $m \le
-1$. It is interesting to point out that the interchange of the
components of the pseudospinors can also be accompanied by some some
relative phases between them.  
\section{The ground state spectrum of a more realistic magnetic field}
\label{realb}
Till now the present article discussed about a magnetic field profile
which becomes infinitely large at around $r=0$. The advantage of
choosing such a magnetic field is related to the fact that the
Dirac-like excitations in graphene in the presence of such a field
becomes localized and the spectrum of the excitations can be exactly
solved. The exact solutions of the previous case could be found
because one can match the problem of the graphene excitations in
presence of an inhomogeneous magnetic field to the exact solutions of
a 2D hydrogen atom bound state spectrum. In this section we will
choose a magnetic field which produces a cut-off Coulomb scalar
potential \cite{sinha,patil} in the effective Schr\"{o}dinger equation
for the upper component $\chi_1$. In this section we will mainly focus
on the localized ground states of the Dirac excitations which can be
exactly found.

If one chooses the function $f(r)$ as
\beq
f(r)=\displaystyle\f{1}{a(r+a)}\,,~~a>0\,,
\label{fr} 
\eeq
where $a$ is a positive constant whose actual dimension is that of
length then the magnetic field  is given by
\beq
B_z(r)=-\f{1}{a(r+a)}-\f{1}{(r+a)^2}\,.
\label{magnetic1}
\eeq
In normal units the magnetic field will be obtained by multiplying
the above expression by $\hbar/e$.  Clearly the magnetic field is a
variable one and finite everywhere. In fact, it lies in the interval
$-{2}/{a^2}\leq B_z(r)\leq 0$. Let us now consider the sector $m\geq
0$. For the above choice of $f(r)$ the scalar potentials $V_{1,2}(r)$
are given by
\beqa
V_1(r)&=&-\f{2m+3}{a(r+a)}+\f{1}{a^2}\,,
\label{v1}\\ 
V_2(r)&=&-\f{2m+3}{a(r+a)}+\f{2}{(r+a)^2}+\f{1}{a^2}\,. 
\label{v2}
\eeqa 
In the present case the two scalar potentials are different.  It may
be noted that apart from the constant $1/a^2$ which may be absorbed in
the energy, $V_1(r)$ is the cut-off Coulomb potential albeit, in two
dimensions. The other potential, $V_2(r)$, is not a cut-off Coulomb
potential since it contains an additional term.  We would like to
point out that the cut-off Coulomb potential of the form
$\displaystyle V(r)=-{\lambda}/(r+a)$, where $\lam>0$, has been
extensively studied. 
The approximate solutions \cite{sinha} as well as
numerical solutions \cite{patil} of the Schr\"{o}dinger equation in an
cut-off Coulomb potential have been found using various
methods. However in most of these studies the potential is treated in
three dimensions where the angular momentum takes positive integer
values starting from zero. On the other hand in the present case we
are working in two dimensions and the coupling depends on $m$. Thus
for $m\geq -1$ the coupling is positive and the potential is an
attractive one. On the other hand for $m<-1$ the potential becomes
repulsive.

We note that Eq.(\ref{chi11}) admits a $\eps=0$ solution and then it is
given by 
\beq
\chi_1=r^{m + 1/2}(r+a)e^{-r/a}\,,
\label{chi1} 
\eeq 
up to a normalization constant.  Clearly $\chi_1(r)\rightarrow 0$ as
$r\rightarrow \infty$, consequently $\chi_1(r)$ is normalizable. 

\begin{figure}[h]
\begin{center}\leavevmode
\includegraphics[width=95mm]{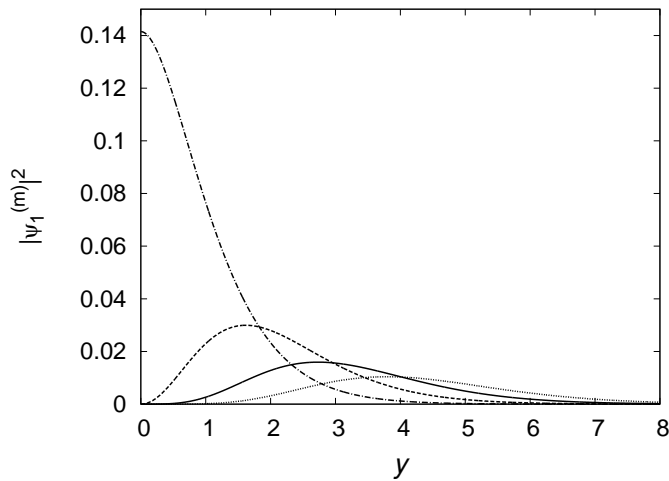}
\caption{Plots of the normalized probability density of the zero-energy
states vs $y$ for different values of $m = 0$ (long dot-dashed),
$ m = 1 $ (dashed), $ m = 2$ (solid) and $m = 3$ (short dot-dashed).
Here $\psi_1^{(m)}$ is the upper component of the pseudospinor for
arbitrary values of $m$.}
\label{Fig6}
\end{center}
\end{figure}

The lower component of the pseudospinor solution of the Dirac equation
can be written down once Eq.(\ref{chi22}) is solved.  The $\eps=0$
solution of Eq.(\ref{chi22}) is given by
\beq
\chi_2(r)=r^{-m - 1/2}\f{e^{r/a}}{r+a}\,,
\label{chi2} 
\eeq 
which is clearly non-normalizable for $m\geq 0$. Therefore the $E=0$
solution of the original problem is given by 
\beq
\psi_{-}^{(m)} (r,\theta)=N_{-}\left(\ba{c}
r^m e^{im\theta}(r+a)e^{-r/a}\\ 0\\ \ea \right),
\eeq 
where $ m = 0,1,2,\cdot\cdot\cdot$.  The normalization constant is 
\beq
N_{-}=\sqrt{\f{2^{2m+3}}{2\pi \,a^{2m+4}(m+3)(2m+3)\Gamma(2m+2)}}\,. 
\eeq 
It may be noted that $\psi_{-}^{(m)}(r,\theta)$ is normalizable for
all values of the quantum number $m\geq 0$. This means that the
$\eps=0$ state i.e, $E=0$ state is infinitely degenerate. Furthermore,
there is no pseudospin down state corresponding to this level.  For
the case where $m<0$, neither the solutions in Eq.~(\ref{chi1}) nor
Eq.~(\ref{chi2}) are normalizable. Consequently for $m<0$ all the
states have non-zero energy. The probability density of the
zero-energy excitations for various angular momentum values are
plotted in Fig.~\ref{Fig6}. The curves show the localized nature of
the states.
\subsection{Reversing the field direction}
For the sake of completeness one can also analyze the spectrum of the
Dirac electron-like excitations of graphene in the specific magnetic
field as given in Eq.~(\ref{magnetic1}) with its sign reversed.  In
this case the vector potentials and the magnetic field are given by
\beq
A_x=-yf(r),~~~~A_y=xf(r)\,,
\label{vector}
\eeq
where $f(r)$ is as given in Eq.~(\ref{fr}).  The zero-energy solutions
of Eq.~(\ref{chi11}) and Eq.~(\ref{chi22}) with the appropriate gauge
potentials as given in Eq.~(\ref{vector}) are given as
\beq
\chi_1= \f{r^{m+1/2}}{r+a}e^{r/a},~~~~\chi_2= 
r^{-(m+1/2)}(r+a)e^{-r/a}\,,
\eeq
up to a normalization constant.  From the above expression it is seen
that $\chi_1$ is not normalizable for any value of $m$. On the other
hand, $\chi_2$ is also not normalizable for $m\geq 0$. However, for
$m<0$ the term $-(m+\f{1}{2})$ becomes positive and consequently
$\chi_2$ becomes normalizable for $m=-1,-2,....$. In other words when
the direction of the magnetic field is reversed, the $E=0$ pseudospin
down states are infinitely degenerate and there are no $E=0$
pseudospin up states. Explicitly the $E=0$ solutions are given by
\beq
\psi_{+}^{(m)}(r,\theta)=N_{+}\left(\ba{c} 0\\
r^{-(m+1)}e^{i(m+1)\theta}(r+a)e^{-r/a}\\
\ea
\right),\\
\label{negb}
\eeq
where $m = -1,-2,-3 \cdot \cdot \cdot$ and $N_{+}$ is obtained from 
$N_{-}$ if one replaces the $m$ by $-(m+1)$. Except these replacement
in $N_{-}$ to obtain $N_{+}$ there can be a possible phase factor
arising out of the magnetic field reversal about which we will discuss
later.  
The important thing to note is that when the magnetic field is
reversed instead of the pseudospin up state we only have the
pseudospin down state in the zero-energy solution. 

In this case also it can be noted that if one reverses the magnetic
field and replaces $m$ by $-(m+1)$ (for a negative $m$ where $m \le
-1$) then $\chi_1$ and $\chi_2$ just gets interchanged. The result of
the field reversal essentially remains the same as discussed in the
previous section. Magnetic field reversal amounts to interchanging the
components of the pseudospinors previously obtained for the magnetic
field pointing in the negative $z$-axis with their positive $m$
values replaced by $-(m+1)$.  Presently $m \le -1$ and the interchange
of the pseudospinor components can be accompanied by a possible
relative phase factor multiplying any of the components.

We would like to note that many of these unusual spectral properties
for 2D inverse-radial magnetic fields was also observed in the
case of planar Schr\"odinger equation in the presence of a
non-constant magnetic field \cite{grosse}. In the present case the
presence of pseudospin adds additional terms to the effective potential and
makes the situation more interesting.
\section{Discussion and conclusion}
\label{conclu}
The present article discusses about the possibility of localizing the
massless Dirac excitations in graphene in the presence of some
specific magnetic field profiles.  The unique structure of the
minimally coupled Dirac equation for a massless particle in two
dimensions gives an interesting physical and mathematical arena to
look out for localized excitations. Localization of the excitations in
graphene is an important topic as conventional electrostatic methods
to localize excited modes fail. The inverse-radial magnetic field, which is
directed along the $z$ direction, has the potential to localize the
graphene excitations near the origin. While there were many previous
works dealing with the topic of magnetic confinement of the
excitations in graphene, most of those works used magnetic fields
which vanished sharply at some radial distance from the origin. The
inverse-radial magnetic field defies this trend of choosing magnetic field
profiles which can produce magnetic confinement of the excitations in
the sense that the inverse-radial magnetic field does not sharply fall or
rise at any distance from the origin. Although the magnetic field
gradually falls away from the origin most of the excitations are
localized near the origin.

It was pointed out in the article that in presence of the inverse-radial
magnetic field the upper component of the pseudospinor of the graphene
excitations follow an effective Schr\"{o}dinger equation. An
interesting observation related to the effective Schr\"{o}dinger
equation of the upper component is related to the fact that this
equation turns out to be the Schr\"{o}dinger equation of the 2D 
hydrogen atom. Consequently if one knows the bound state
spectrum of the 2D Hydrogen atom one can also solve for
the upper component of the graphene excitations in presence of an
inverse-radial magnetic field. The lower component of the spinor can easily
be found by the application of the Dirac equation. Our work gives a
natural and physical application to the 2D hydrogen atom
problem which by itself remains an esoteric and formal problem to deal with. 

In the case of the finite magnetic field the relevant Dirac equation
for the excitations had a close connection with the 2D cut-off
Coulomb potential. The Schr\"{o}dinger equation for a cut-off Coulomb
potential still remains to be analytically solved. Consequently for
the realistic magnetic field the full excitation spectrum could not be
presented because to present the full spectrum one has to solve an
effective Schr\"{o}dinger equation in presence of the 2D cut-off
Coulomb potential. 

One of the interesting features of the spectra of the graphene
excitations in presence of the inhomogeneous magnetic fields is
related to the degeneracy pattern of the solutions. We have tried to
point out some of the relevant degeneracy patterns and a general way
how these degeneracies arise in the case of the inverse-radial
magnetic field.  The ground state of the magnetically localized ground
state turns out to be infinitely degenerate in both the cases. This is
valid when the system size is infinite and the wave function vanishes
at infinity. In principle, infinite number of electrons can be
accommodated in the zero-energy states. For a realistic situation the
degeneracy in the energy levels can not be infinity because of the
finite size of the system. Fig. \ref{Fig3} shows that the position of
the peak of the densities for various cases is shifting away from the
origin as we increase the quantum numbers $(n,m)$. The distance of the
point where the density maximizes can not be greater than the radius
of the system. This implies that there is an upper cut-off in the
$(n,m)$ values. Therefore, the degeneracy is finite in a realistic
situation.

Throughout our analysis we have tried to see the effects of a magnetic
field direction reversal on the excitation spectrum. The result
obtained shows up most prominently in the zero-energy states. When the
magnetic field is pointed in the negative $z$-direction only the upper
component of the pseudospinor survives where as reversing the field
direction kills the upper component and only the lower component of
the pseudospinor exists. In the most general case it was pointed out
that a magnetic field reversal interchanges the pseudospinor
components accompanied by a change of the $m$ quantum numbers to
$-(m+1)$ where now $m \le -1$. It was also pointed out for both the
cases, discussed in this article, such an interchange of the
pseudospinor components could be accompanied by a relative phase
arising out of the field reversal. In this section we can shed some
light on the relative phase accompanying such a magnetic field
reversal.

It is interesting to connect these phenomenon related to
the magnetic field reversal with the time-reversal like operation
${\cal S}$ as discussed in the review by Beenakker \cite{beena}. Under
the action of ${\cal S}$ the Hamiltonian of the excitation in
Eq.~(\ref{bham}) changes as
\begin{eqnarray}
{\cal S} H({\bf A}) {\cal S}^{-1}=H(-{\bf A})\,,
\end{eqnarray}
where ${\bf A}$ is the gauge potential giving rise to the magnetic
field. Although ${\cal S}$ is not the exact time-reversal operator,
the exact time reversal operator changes the valleys, it acts like the
time-reversal operator in a single valley. The action of ${\cal S}$ is
like reversing the magnetic field direction. The operator ${\cal
  S}=i\sigma_2 C$ where $C$ is the complex conjugation operation. It
can easily be verified that if one apply this operator on the
pseudospinors one gets the pseudospinors of the reversed field case,
as discussed in this article. As a result of the ${\cal S}$ operation
on the pseudospinors the component of the previous pseudospinor
components gets interchanged with a relative sign change. This
appearance of a sign can be easily accommodated in the scheme discussed
previously where it was pointed out that the interchange of the
pseudospinor components can be accompanied by a relative phase.

To conclude this article it can be stated that the present article
tries to analyze whether one can have magnetically confined
excitations in graphene with inhomogeneous magnetic fields which do
not have any finite/infinite discontinuities in the non-zero finite
range of the radial coordinate. The magnetic field profiles chosen
turns out to be interesting because their specific functional forms
allows one to use the results obtained for the bound states of 2D
Schr\"{o}dinger equations in the presence of a Coulomb or a cut-off
Coulomb potential. It is observed that there exists an infinite number
of localized excitations which are magnetically confined.  In the
future if one can produce a inverse-radial magnetic field then these exotic
localized excitations can be observed experimentally.
\begin{acknowledgements}
We would like to thank Sudeep Bhattacharjee for a discussion on
realization of the inverse-radial magnetic field. 
\end{acknowledgements}

\end{document}